\documentclass[10pt,
final,
 conference, 
twocolumn,
comsoc,
 letterpaper]{IEEEtran} 
\usepackage{srcltx}
\usepackage[psamsfonts]{amsfonts}
\usepackage{amsmath}
\usepackage{amssymb}
\usepackage{graphics}
\usepackage{psfrag}
\usepackage{epsfig}
\usepackage{algorithm}
\usepackage{algorithmic}
\usepackage{array}
\usepackage{mathtools}
\usepackage[psamsfonts]{amsfonts}
\usepackage{makeidx}  
\usepackage{bbm}
\usepackage{amsfonts}
\usepackage{hhline}
\usepackage{eufrak}
\usepackage{yfonts}
\usepackage{pifont}
\usepackage{footnote}
\usepackage{pdfpages}
\usepackage{graphicx}
\usepackage{balance}
\usepackage{color}
\usepackage[square, comma, sort&compress, numbers]{natbib}
\usepackage{dsfont}
\usepackage{caption}
\usepackage{subcaption}
\usepackage{sidecap}
\usepackage[splitrule,hang,marginal]{footmisc}
\usepackage{enumitem}
\usepackage{bm}
\usepackage{tikz}
\usepackage{comment}
\usetikzlibrary{fit,positioning}

\DeclarePairedDelimiter\floor{\lfloor}{\rfloor}

\newtheorem{cor}{Corollary}

\IEEEoverridecommandlockouts

\makeatletter
\newcommand\fs@spaceruled{\def\@fs@cfont{\bfseries}\let\@fs@capt\floatc@ruled
  \def\@fs@pre{\vspace{0.9\baselineskip}\hrule height.8pt depth0pt \kern2pt}%
  \def\@fs@post{\kern5pt\hrule\relax}%
  \def\@fs@mid{\kern5pt\hrule\kern2pt}%
  \let\@fs@iftopcapt\iftrue}
\makeatother

\newcommand{\NR}[1]{{\color{orange} NR: #1}}
\newcommand{\MJ}[1]{{\color{blue} MJ: #1}}

\title{Primary User Localization and Online Radio Cartography via Structured Tensor Decomposition   
\author{Mohsen Joneidi and Nazanin Rahnavard\\
Department of Electrical and Computer Engineering, University of Central Florida\\
\{joneidi, nazanin\}@eecs.ucf.edu
\vspace{-3mm}
\thanks{This material is based upon work supported by the National Science Foundation under Grant No. CCF-1718195.}}}
\begin{document}

\maketitle

\begin{abstract} 
Source localization and radio cartography using multi-way representation of spectrum is the subject of study in this paper. 
A joint matrix factorization and tensor decomposition problem is proposed and solved using an iterative algorithm.
The multi-way measured spectrum is organized in a tensor and it is modeled by multiplication of a propagation tensor and a channel gain matrix.
The tensor indicates the propagating power from each location and each frequency over time and the channel matrix links the propagating tensor to the sensed spectrum. 
We utilize sparsity and other intrinsic characteristics of spectrum to identify the solution of the proposed  problem. Moreover, The online implementation of the proposed framework  results in online radio cartography which is a powerful tool for efficient spectrum awareness and utilization. The simulation results show that our algorithm is a promising technique for dynamic primary user localization and online radio cartography.  
\end{abstract}
\begin{IEEEkeywords}
Radio cartography,  tensor CP decomposition, source localization, and structural sparsity. 
\end{IEEEkeywords}
\vspace{-1.5mm}
\section{Introduction}
High dimensional analysis of sensed spectrum data  is an enabling step towards the efficient utilization of spectrum  \cite{glandon2017recurrent, lu2017wideband}. Primary users (PUs) of licensed spectrum often under-utilize this valuable resource \cite{Akyildiz06_xG}. In the cognitive radio network paradigm, the unlicensed or secondary users (SUs) are allowed to coexist with PUs if they do not interfere with PUs~\cite{8240657}.

 Tensor decomposition and multi-way modeling of data  are old problems in mathematics \cite{cattell1952three}.  CANDECOMP/PARFAC (CP) decomposition  and Tucker decomposition  are two well-known tensor decomposition methods,  which can be interpreted as two extensions of matrix singular value decomposition \cite{carroll1970analysis,tucker1966some}. Tensor-based methods have been employed in communications and coding frameworks since Sidiropoulos et. al. introduced them for blind code-division multiple access  (CDMA) \cite{sidiropoulos2000blind}. 
Recently, more complex communication systems are developed based on the tensor decomposition theory \cite{da2018tensor,boutalline2015blind,sorensen2017blind}. The common idea of all the tensor-based coding methods is to perform a joint blind symbol and channel estimation at the receiver, which is feasible due to mild conditions for uniqueness of tensor CP decomposition \cite{favier2012tensor}. In this paper, factorization of the sensed tensor to a latent tensor (propagation tensor) and a  transformation matrix (perturbed channel gains) is studied. This problem can also be seen as a generalization of three-way compressed sensing \cite{Sidiropoulos12TensorCS}, when the compression (sensing matrix) is not fully determined. 

In this paper, the received spectrum power at SUs, also referred to as spectrum sensors, is organized as a 3-dimensional tensor, called the \emph{sensed tensor}. Moreover, a latent tensor corresponding to the spatial, spectral and temporal power propagation patterns of the PUs is defined and referred to as the \emph{propagation tensor}. The goal is to estimate the propagation tensor via structured decomposition of the given sensed tensor. The compression operator between the sensed tensor and the propagation tensor is not fully determined, due to randomness of channel gains. The sensed tensor is an compressed replica of the unknown propagation tensor and the channel gain matrix is the compressor operator. 
There exist some efforts for studying the effect of perturbation on compressive sensing stability \cite{yang2012robustly}. This paper exploits the link between CP components of the sensed tensor and those of the unknown propagation tensor. Estimating the propagation tensor can lead to estimating the RF power spectrum at any arbitrary location of the network  according to some channel propagation model. Generating an RF spectrum map is also known radio cartography. In this paper, we will employ our tensor decomposition algorithm and introduce an online implementation for radio cartography. 

The main contributions of this paper are summarized as follows: 

\begin{itemize}
    \item Multi-way compressive sensing is regularized for the spectrum sensing problem and it is employed for radio cartography. 
    \item Spatial, spectral, and temporal structures are imposed for different dimensions of the tensor accordingly. 
    \item The proposed algorithm is robust against perturbations in the channel gain matrix.
    \item The proposed algorithm  is implemented for online radio cartography where the dynamic of PUs is changing over time.
\end{itemize}


\textbf{Notations:} Throughout this paper, vectors, matrices, and tensors are denoted by bold lowercase, bold uppercase, and bold underlined uppercase letters, respectively. 
A fiber is defined by fixing every index of a tensor but one.
For example, for $\underline{\boldsymbol{T}} \in \mathbb{R}^{N \times M \times K}$, $\underline{\boldsymbol{T}}_{:,j,k} $ is a vector of length $N$, also known as the mod-1 fiber of $\underline{\boldsymbol{T}}$.  $\boldsymbol{T}_1$, $\boldsymbol{T}_2$, and $\boldsymbol{T}_3$ are unfolded matrices whose columns are fibers of the first, second and third dimensions of $\underline{\boldsymbol{T}}$, respectively. Slices are two-dimensional sections of a tensor, defined by fixing all but two
indices. For a third-order tensor $\underline{\boldsymbol{T}}$,  $\underline{\boldsymbol{T}}_{:,:,k} $ is an $N \times M$ matrix, which is also called the $k^{th}$ frontal slice $\underline{\boldsymbol{T}}$. Horizontal and lateral Slices of $\underline{\boldsymbol{T}}$ are denoted by $\underline{\boldsymbol{T}}_{n,:,:} $ and  $\underline{\boldsymbol{T}}_{:,m,:}$, respectively.
 Khatri-Rao product is denoted by $\odot$. 
Moreover, $\circ$ denotes the outer product.
 The n-mode product of a tensor $\boldsymbol{\underline{X}}$ with a proper sized transformation matrix $\boldsymbol{U}$ is a tensor and denoted by $\boldsymbol{\underline{X}}\times_n \boldsymbol{U}$. It transfers each fiber of the $n^{\text{th}}$ mode of tensor to the corresponding fiber in the final tensor. Mathematically,
$$
\boldsymbol{\underline{Y}}=\boldsymbol{\underline{X}}\times_n \boldsymbol{U} \leftrightarrow \boldsymbol{Y}_n=\boldsymbol{U}\boldsymbol{X}_n, 
$$
in which $\boldsymbol{X}_n$ and $\boldsymbol{Y}_n$  are unfolded replicas of tensor $\boldsymbol{\underline{X}}$ and $\boldsymbol{\underline{Y}}$ w.r.t. different dimensions.

\section{System Model and the CP Decomposition}\label{sec2}
The system model in this paper is similar to that of~\cite{Bazerque10Distributed}. Suppose an area of interest is divided into $P$ grid points  and transmitters (PUs) are assumed to be located in a subset of these grid points, unknown to us. On the other hand, there are $N$ receivers (SUs) with known locations that receive a superposition of the transmitters' signals. The received signal is affected by the channel gains $\gamma_{pn}$ between the transmitters and receivers and observed in the presence of zero mean AWGN with variance $\sigma_n^2$.  Due to simplicity and similar to \cite{Bazerque10Distributed,Bazerque11GroupLasso}, we adopt a pathloss model, i.e., the received power spectral density (PSD) at a sensor is proportional to the reciprocal of distance between the location of transmitter and the receiver. Since only the received power at each frequency is modeled and phase information is neglected, spectrum sensors are not required to be synchronized. Moreover, multi-path effect and the phase delay of each path are neglected as well as phase delay of the line-of-sight path.  
Accordingly, $\gamma_{pn}$, the channel gain between location $p$ and sensor $n$ is defined by $1/d_{pn}^\eta$, where $d_{pn}$ indicates the spatial distance between the grid point $p$ and receiver $n$ and $\eta$ is the loss factor of environment which is a constant between 2 and 3 \cite{Bazerque10Distributed}.
The received PSD at receiver/sensor $n$ at time $t$ and frequency bin $f$ can be written as,
\begin{equation}\label{rec2}
\small
\begin{split}
y_n (t,f)=\sum_{p=1}^{P}\gamma_{pn} x_p(t,f)+v_n(t,f)
=\boldsymbol{\gamma}_n^T \boldsymbol{x}(t,f) +v_n(t,f),
\end{split}
\end{equation}
where, $\small{\boldsymbol{\gamma}_n^T =[\boldsymbol{\gamma}_{1n} \;\boldsymbol{\gamma}_{2n} \;\ldots\;\boldsymbol{\gamma}_{P n}]}$ contains the channel gains between all grid locations and the $n^{\text{th}}$ sensor. Moreover, $x_p(t,f)$ denotes the propagation power at the grid point $p$ at time $t$ and frequency bin $f$ and $\boldsymbol{x}(t,f)=[x_1(t,f)\; x_2(t,f)\ldots \;x_P(t,f)]^T$. Further, $v_n(t,f)$ is the received noise at the $n^{\text{th}}$ sensor at time $t$ and frequency bin $f$. We assume the measurements are available for $T$ time slots and $F$ frequency bins. Collaborative estimation of $\boldsymbol{x}(t,f)\in \mathbb{R}^{P}$ over each time slot  and for each frequency bin requires collecting measurements of all sensors in vector $\boldsymbol{y}(t,f)\in \mathbb{R}^{N}$. The following  minimization problem has been proposed for the estimation of $\boldsymbol{x}(t,f)$ for each time and frequency independently~\cite{Bazerque10Distributed,8108778},
\begin{equation}
    \label{eq:l1}
\boldsymbol{x} (t,f)=\underset{\boldsymbol{x}}{\text{argmin}}\; \|\boldsymbol{y} (t,f)-\boldsymbol{\Gamma}\boldsymbol{x}\|_2,
\end{equation}
where the $n^{\text{th}}$ row of matrix $\boldsymbol{\Gamma}$ is  $\boldsymbol{\gamma}_n^T$. 
 Problem (\ref{eq:l1}), subjected to $\ell_1$  constraint on $\boldsymbol{x}$, has previously been employed for collaborative spectrum estimation for a given $\boldsymbol{\Gamma}$ \cite{Bazerque10Distributed,8108778}. However, in our work, the sensed data is organized as a tensor, which facilitates exploiting data structures in spatial, spectral, and temporal domains.
 
\begin{figure}[h]
\centering
\includegraphics[width=2.5in,height=3.2cm]{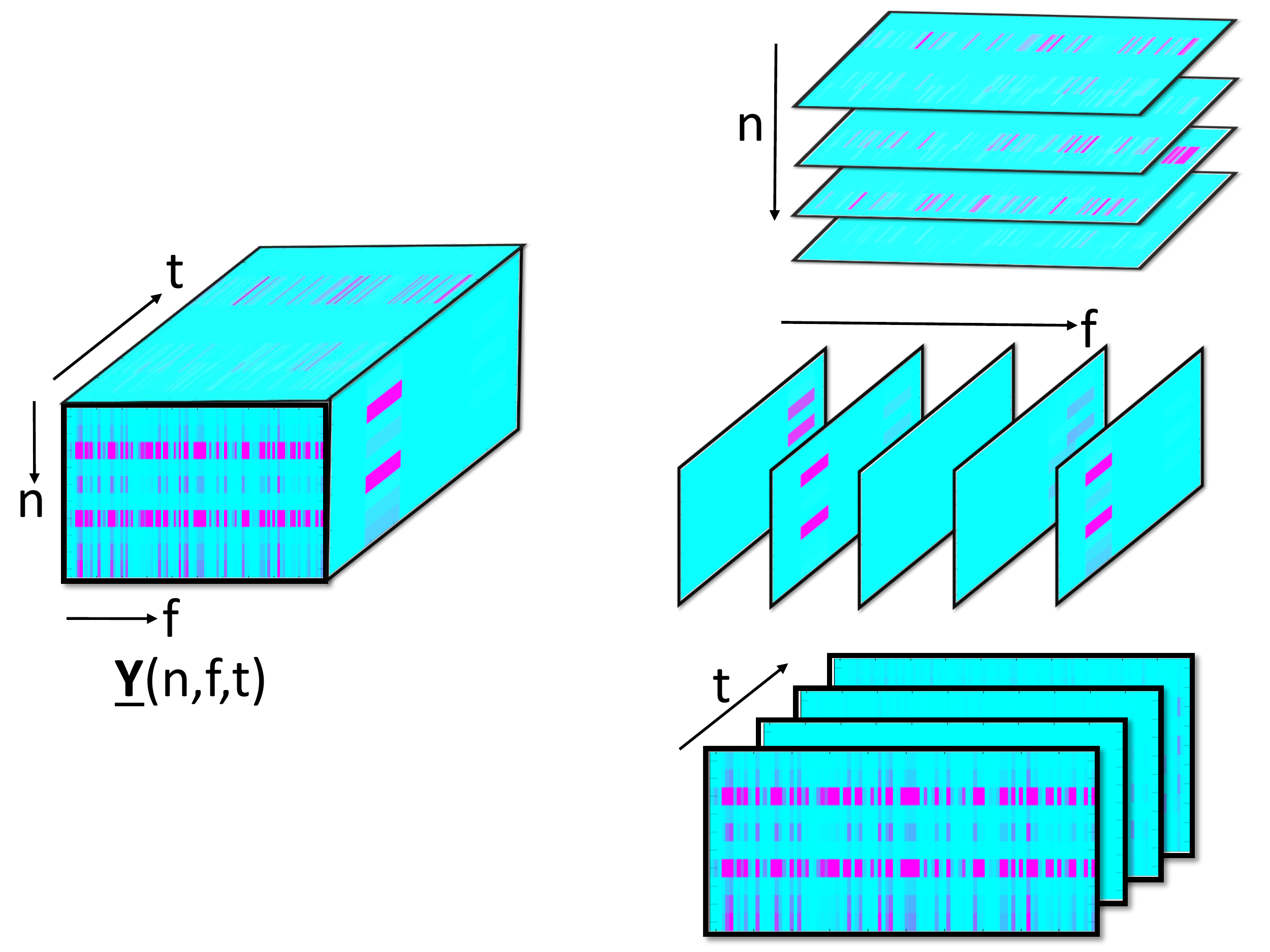}
\caption{\small Structure of tensor $\underline{\boldsymbol{Y}} \in \mathbb{R} ^{N \times F \times T}$, and its slices in time, frequency, and space.}\label{fig:power_tensor}
\end{figure}

 Received spectrum power at sensor $n$ is stored at the $n^{\text{th}}$ slice of the first way of tensor $\underline{\boldsymbol{Y}}$. Likewise, received spectrum power at frequency $f$  is stored at the $f^{\text{th}}$ slice of the second way and received data at time $t$ are placed in the  $t^{\text{th}}$ slice of the third way of tensor 
$\underline{\boldsymbol{Y}}$. Fig. \ref{fig:power_tensor} shows the structure of the sensed tensor $\underline{\boldsymbol{Y}}$. Tensor $\underline{\boldsymbol{X}}$ also can be defined where $x_{pft}$ indicates power propagation from location $p$ in frequency bin $f$ at time $t$. Having only access to a compressed sensed tensor $\underline{\boldsymbol{Y}}$, we aim at decomposing $\underline{\boldsymbol{X}}$ in order to reveal propagation pattern of the network. 

Our proposed tensor-based approach is mainly based on CP decomposition~\cite{kolda2009tensor}, which factorizes a tensor into a sum of rank-one tensors. For example, a three-way tensor $\underline{\boldsymbol{X}}\in \mathbb{R}^{P\times F\times T}$ can be decomposed to summation of $R$ rank-1 tensors as follows,
\begin{equation}
\small
\underline{\boldsymbol{X}}=\sum_{i=1}^R   \boldsymbol{a}_i \circ \boldsymbol{b}_i \circ \boldsymbol{c}_i,
\label{CP_def}
\end{equation}
where, $\boldsymbol{a}_r\in \mathbb{R}^P$, $\boldsymbol{b}_r\in \mathbb{R}^F$ and $\boldsymbol{c}_r\in \mathbb{R}^T$ are \emph{factor vectors} of the $r^{\text{th}}$ rank-one component. Moreover, $R$ is referred as the rank of CP decomposition. The \emph{factor matrices} refer to the collection of vectors from the rank-one components, i.e, $\boldsymbol{A}=[\boldsymbol{a}_1 \boldsymbol{a}_2 \ldots \boldsymbol{a}_R]\in \mathbb{R}^{P\times R}$ and likewise for $\boldsymbol{B}\in \mathbb{R}^{F\times R}$ and $\boldsymbol{C}\in \mathbb{R}^{T \times R}$. As the inherent structures of data are preserved in tensor representation, we can easily exploit prior knowledge on the CP factors of a tensor. The constraints imposed by the dynamics of the problem result in achieving more accurate analysis in a concise mathematical formulation. The alternating least squares algorithm is a well-known method for obtaining CP factors \cite{kolda2009tensor}. 

\section{Tensor Formulation of Spectrum Sensing}
\label{sec:pr}
Large tensor processing may require compression of all dimensions of tensor in order to make the problem computationally tractable \cite{sidiropoulos2014parallel,papalexakis2015p}. However, in some practical scenarios, the nature of a problem implies compression in one or more dimensions of the tensor of interest. For example, in our spectrum sensing framework, we have access to a linearly transformed replica of the unknown tensor $\underline{\boldsymbol{X}}$. Mathematically speaking, we have 
\begin{equation}
\small
\label{main_problem}
\begin{split}
&\underline{\boldsymbol{Y}}=\underline{\boldsymbol{X}}\times_1 (\boldsymbol{\Gamma}_M+\boldsymbol{\Gamma}_p)+\underline{\boldsymbol{N}},\\
&\underline{\boldsymbol{X}}=\sum_{r=1}^R   \boldsymbol{a}_r \circ \boldsymbol{b}_r \circ \boldsymbol{c}_r,
\end{split}
\end{equation}
in which,  $\boldsymbol{\Gamma}_M$ is the known compressor matrix imposed by the underlying model in (\ref{rec2}) and $\boldsymbol{\Gamma}_p$ is the unknown perturbation (it models the discrepancy between the assumed and actual gains).   $\boldsymbol{a}_r$, $ \boldsymbol{b}_r $ and $ \boldsymbol{c}_r $ indicate the $r^{\text{th}}$ factor vectors of the big tensor $\underline{\boldsymbol{X}}$ while $\underline{\boldsymbol{N}}$ models the additive measurement noise. Each factor follows a certain structure which originates from the nature of spectrum data. For instance, spatial and spectral activation patterns are sparse due to the scarce presence of PUs in the area of interest and  their relatively narrow bands of communication, respectively. 

To obtain the latent factors of the tensor $\underline{\boldsymbol{X}}$ and to find the perturbation $\boldsymbol{\Gamma}_{P}$, spatial factors (columns of $\boldsymbol{A}$) are considered to be 1-sparse vectors which implies that the activity pattern of each spatial point is captured via a rank-1 tensor. In other words, the propagation of each location is considered independent of other locations, i.e., each grid location has a certain spectral and temporal activation pattern. The goal is to find these patterns for each location on the grid. 
Summation of $\boldsymbol{a}_r$'s is an $R$-sparse vector whose support corresponds to the locations of power propagation for the whole area from maximum $R$ locations (PUs).  Sparsity for locations of active PUs and sparsity of their spectrum usage is a common assumption~\cite{Bazerque10Distributed}. Total variation of temporal activation factors is bounded according to the previous time sample of the factor.  Piece-wise constancy is a proper assumption for temporal behavior of primary users in a CRN \cite{zaeemzadeh2015missing}. Likewise, each entry of the compressor matrix is bounded around the compressor matrix of the model. With the above assumptions, we propose the following structural CP decomposition as our spectrum sensing problem,
\begin{equation}
\label{compressed_CP}
\begin{split}
\small
\underset{\boldsymbol{a}_r,\boldsymbol{b}_r,\boldsymbol{c}_r,\boldsymbol{\Gamma}_p}{\text{argmin}}&\;\; \|\underline{\boldsymbol{Y}}-\underline{\boldsymbol{X}}\times_1 (\boldsymbol{\Gamma}_M +\boldsymbol{\Gamma}_p)\|_F^2+\lambda_p\|\boldsymbol{\Gamma}_p\|_F^2 \\
&+\lambda_b \sum_r  \|\boldsymbol{b}_r\|_1 +\lambda_c\sum_r \sum_{t=2}^T(c_{tr}-c_{t-1,r})^2\\
\text{subject to:} &\quad 
\underline{\boldsymbol{X}}=\sum_{r=1}^R   \boldsymbol{a}_r \circ \boldsymbol{b}_r \circ \boldsymbol{c}_r \;\;\text{and}\; \|\boldsymbol{a}_r\|_0=1,
\end{split}
\end{equation}
in which, $\lambda_p$, $\lambda_b$ and $\lambda_c$ regularize the problem for imposing the desired aforementioned structures. The number of non-zero entries of a vector is denoted by $\|.\|_0$ and $\|.\|_1$ indicates $\ell_1$ norm of a vector.

From~(\ref{main_problem}), we can find the relation between the factor matrices of $\boldsymbol{X}$ and $\boldsymbol{Y}$. The second and the third sets of factors of $\boldsymbol{X}$ and $\boldsymbol{Y}$,i.e., $\boldsymbol{B}$ and $\boldsymbol{C}$, remain the same as the compression in ~(\ref{main_problem}) only is done in mod-1. The relation between the first set of factors of $\boldsymbol{Y}$ (denoted by $\tilde{\boldsymbol{A}}$) and that of $\boldsymbol{X}$ (denoted by $\boldsymbol{A}$) is given by~\cite{sid2012multi}

\begin{equation}
\label{pert1}
\tilde{\boldsymbol{A}}=(\boldsymbol{\Gamma}_M+\boldsymbol{\Gamma}_p)\boldsymbol{A}.
\end{equation}
In other words, we aim to solve a three-way tensor decomposition where one way of the tensor is compressed and there is an uncertainty about the compression matrix. 

\section{Solving the joint Problem}
\label{sec:sol}

The joint optimization problem for estimation of structured factors and the perturbation matrix is solved via an extension of alternative least squares (ALS) framework \cite{kolda2009tensor}. Let us break down (\ref{compressed_CP}) to four sub-problems as follows,

\small{
\begin{equation}\begin{split}
\hat{\boldsymbol{A}}&=\underset{\boldsymbol{A}}{\text{argmin}} \|\boldsymbol{Y}_1 -\boldsymbol{\Gamma}\boldsymbol{A}(\hat{\boldsymbol{C}}\odot \hat{\boldsymbol{B}})^T\|_F^2\;\;\text{s.t.} \; \|\boldsymbol{a}_r\|_0=1,\;\; \forall r\\
\hat{\boldsymbol{B}}&=\underset{\boldsymbol{B}}{\text{argmin}} \|\boldsymbol{Y}_2 -\boldsymbol{B}(\hat{\boldsymbol{C}}\odot \boldsymbol{\Gamma}\hat{\boldsymbol{A}})^T\|_F^2+\lambda_b \sum_r\|\boldsymbol{b}_{r}\|_1,\\
\hat{\boldsymbol{C}}&=\underset{\boldsymbol{C}}{\text{argmin}} \|\boldsymbol{Y}_3 -\boldsymbol{C}(\hat{\boldsymbol{B}}\odot \boldsymbol{\Gamma}\hat{\boldsymbol{A}})^T\|_F^2+\lambda_c \sum_r\sum_t|c_{tr}-c_{t-1,r}|^2,\\
\hat{\boldsymbol{\Gamma}}_p&=\underset{\boldsymbol{\Gamma}_p}{\text{argmin}} \|\boldsymbol{Y}_1 -(\boldsymbol{\Gamma}_M+\boldsymbol{\Gamma}_p)\hat{\boldsymbol{A}}(\hat{\boldsymbol{C}}\odot \hat{\boldsymbol{B}})^T\|_F^2+\lambda_p\text{tr}(\boldsymbol{\Gamma}_p\boldsymbol{\Gamma}_p^T).
\label{ALS_our}
\end{split}
\end{equation}
}
\normalsize{In the first three problems, $\boldsymbol{\Gamma}=\boldsymbol{\Gamma}_M+\hat{\boldsymbol{\Gamma}}_p$. Sub-problems w.r.t. $\boldsymbol{B}$, $\boldsymbol{C}$ and $\boldsymbol{\Gamma}_p$ are regular regression problems. However, by sparsity assumption on $\boldsymbol{a}_r$, the first sub-problem turns into a \emph{compressive sensing} problem in which $\boldsymbol{\Gamma}_M+\boldsymbol{\Gamma}_p$ can be regarded as a measurement matrix.  However, non-zero entries of spatial factors are distributed among factors of $\boldsymbol{A}$ and each column has only one non-zero entry. Thus, the spatial activation pattern can be easily revealed using a simple compressed sensing reconstruction algorithm. In this paper, we use orthogonal matching pursuit (OMP) algorithm, a low-complexity greedy algorithm for sparse recovery \cite{tropp2007signal}. The problem w.r.t. $\boldsymbol{B}$ and $\boldsymbol{C}$ would be a regularized least squares (LS) problem. However, the solution w.r.t.  $\boldsymbol{B}$ is an $\ell_1$ regularized least squares solution which is known as LASSO~\cite{angelosante2009rls}. The third sub-problem w.r.t. $\boldsymbol{C}$ is a smoothed LS problem. We refer to the solution as SLS(.,.). 
}
In order to solve this iterative problem w.r.t. $\boldsymbol{A}$, first we need to develop a corollary based on Theorem 1 in Sec. 6.9 of \cite{luenberger1997optimization}. 
\begin{cor} 
\label{lem:ls_eq}
Let $\boldsymbol{G}$ and $\boldsymbol{Z}$ denote two matrices with their first dimension is smaller than their second dimension. The following least squares minimization problem w.r.t. $\boldsymbol{X}$, 
$$
\underset{\boldsymbol{X}}{\text{argmin}}\|\boldsymbol{Y}-\boldsymbol{GXZ}\|_F^2 
$$
is equivalent to solving the following equality
$$
\boldsymbol{YZ}^+=\boldsymbol{GX},
$$
where $(\;)^+$ indicates the pseudo-inverse operator.
\end{cor}

According to the aforementioned corollary, the first sub-problem w.r.t. $\boldsymbol{A}$ can be written as
$$
\hat{\boldsymbol{A}}=\underset{\boldsymbol{A}}{\text{argmin}} \|\boldsymbol{Y}_1 (\hat{\boldsymbol{C}}\odot \hat{\boldsymbol{B}})^+ -(\boldsymbol{\Gamma}_M+\hat{\boldsymbol{\Gamma}}_p)\boldsymbol{A}\|_F^2\;\;\text{s.t.}\; \|\boldsymbol{a}_r\|_0=1.
$$
Let us define $\boldsymbol{U}=\boldsymbol{Y}_1 (\hat{\boldsymbol{C}}\odot \hat{\boldsymbol{B}})^+$. The $r^{\text{th}}$ column of $\boldsymbol{U}$ corresponds to the same column in $\boldsymbol{A}$. Thus, the problem w.r.t. $\boldsymbol{a}_r$ for $r=1,2,\ldots,R$ can be cast as 
$$
\hat{\boldsymbol{a}}_r=\underset{\boldsymbol{a}}{\text{argmin}} \|\boldsymbol{u}_r -(\boldsymbol{\Gamma}_M+\hat{\boldsymbol{\Gamma}}_p)\boldsymbol{a}\|_F^2\;\;\text{s.t.}\; \|\boldsymbol{a}_r\|_0=1,
\label{CS_factor}
$$
where, $\boldsymbol{u}_r$ denotes the $r^{\text{th}}$ column of $\boldsymbol{U}$. We refer to the solution of this problem as OMP($\boldsymbol{u}_r$,$\boldsymbol{\Gamma}_M+\hat{\boldsymbol{\Gamma}}_p,1$). 
Accordingly, $\hat{\boldsymbol{B}}$ and $\hat{\boldsymbol{C}}$ are optimized by solving LASSO and the SLS problem, respectively. The subproblem w.r.t $\boldsymbol{\Gamma}_p$ has a closed-form solution which can be derived by taking derivative w.r.t. $\boldsymbol{\Gamma}_p$. The update rule can be easily derived as 
\begin{equation}
\label{eq:cl_gamma}
\boldsymbol{\Gamma}_p=\boldsymbol{E}_1 \boldsymbol{X}_1^T (\boldsymbol{X}_1\boldsymbol{X}_1^T+\lambda_p \boldsymbol{I})^{-1},
\end{equation}
where, {\small$\underline{\boldsymbol{E}}=\underline{\boldsymbol{Y}}-\underline{\boldsymbol{X}}\times_1\Gamma_M$}. The solution of (\ref{eq:cl_gamma}) is referred as the regularized least squares (RLS). Steps of the proposed  alternating constrained LS algorithm for solving the joint  problem of propagation tensor estimation and perturbation matrix updating is summarized in Alg. \ref{alg:als1}. 


\floatstyle{spaceruled}
\restylefloat{algorithm}

\begin{algorithm}[t!]
\caption{Alternating constrained least squares algorithm for solving the joint tensor decomposition and matrix factorization problem in ~(\ref{ALS_our})}
\label{alg:nn}
\begin{algorithmic}[1]
\REQUIRE Tensor $\underline{\boldsymbol{Y}}$, $\lambda_p$, $\lambda_b$, $\lambda_c$, $\boldsymbol{\Gamma}_M$ and $R$. \\ 
\hspace{-4.3mm}
\textbf{Output:} $\boldsymbol{A}, \boldsymbol{B}, \boldsymbol{C}$ and $\boldsymbol{\Gamma}_P$.\\
\hspace{-4.3mm}
\textbf{Initialize:} \text{$\tilde{\boldsymbol{A}}$, $\boldsymbol{B}$ and $\boldsymbol{C}$ using conventional CPD \cite{kolda2009tensor} }
\\ 
\hspace{-4.3mm}
\textbf{Initialize:} $\boldsymbol{\Gamma}_P=\boldsymbol{0}$\\ 

\textbf{While} ({The stopping criterion is not met})\\
  $\quad \boldsymbol{A}\leftarrow\;\text{OMP}(\boldsymbol{U},\boldsymbol{\Gamma}_M+\boldsymbol{\Gamma}_p,1)$\\
  $\quad\boldsymbol{B}\leftarrow\;\text{LASSO}(\boldsymbol{Y}_2^T,\boldsymbol{C}\odot (\boldsymbol{\Gamma}_M+\boldsymbol{\Gamma}_p){\boldsymbol{A}},\lambda_b)$\\
   $\quad\boldsymbol{C}\leftarrow\;\text{SLS}(\boldsymbol{Y}_3^T,{\boldsymbol{B}}\odot (\boldsymbol{\Gamma}_M+\boldsymbol{\Gamma}_p){\boldsymbol{A}},\lambda_c)$\\
   $\quad \boldsymbol{\Gamma}\leftarrow\;\boldsymbol{\Gamma}_M+ \; \text{RLS}(\boldsymbol{E}_1,\boldsymbol{X}_1,\lambda_p)$
\end{algorithmic}
\label{alg:als1}
\end{algorithm}

Estimating the propagation tensor is sufficient to interpolate spectrum at any arbitrary location. The reconstructed spectrum at any point of the network, $\boldsymbol{g}$, is called spectrum map and can be computed as follows:
\begin{equation}
\small
M(\boldsymbol{g},f,t)=\sum_p \sum_r \frac{a_r(p)b_r(f)c_r(t)}{\text{dist}(\boldsymbol{g},\text{grid}(p))^{\eta}},
\label{eq:power_map}
\end{equation}
in which, dist(,) corresponds to the Euclidean distance between its arguments and $\text{grid}(p)$ is the location of $p^{\text{th}}$ grid point. Radio cartography for different frequencies and any location of network, $M(\boldsymbol{g},f,t)$, is the final output of our proposed framework.  

\section{Online Implementation}
\begin{figure}[h]
\centering
\includegraphics[width=3.45in,height=2.35in]{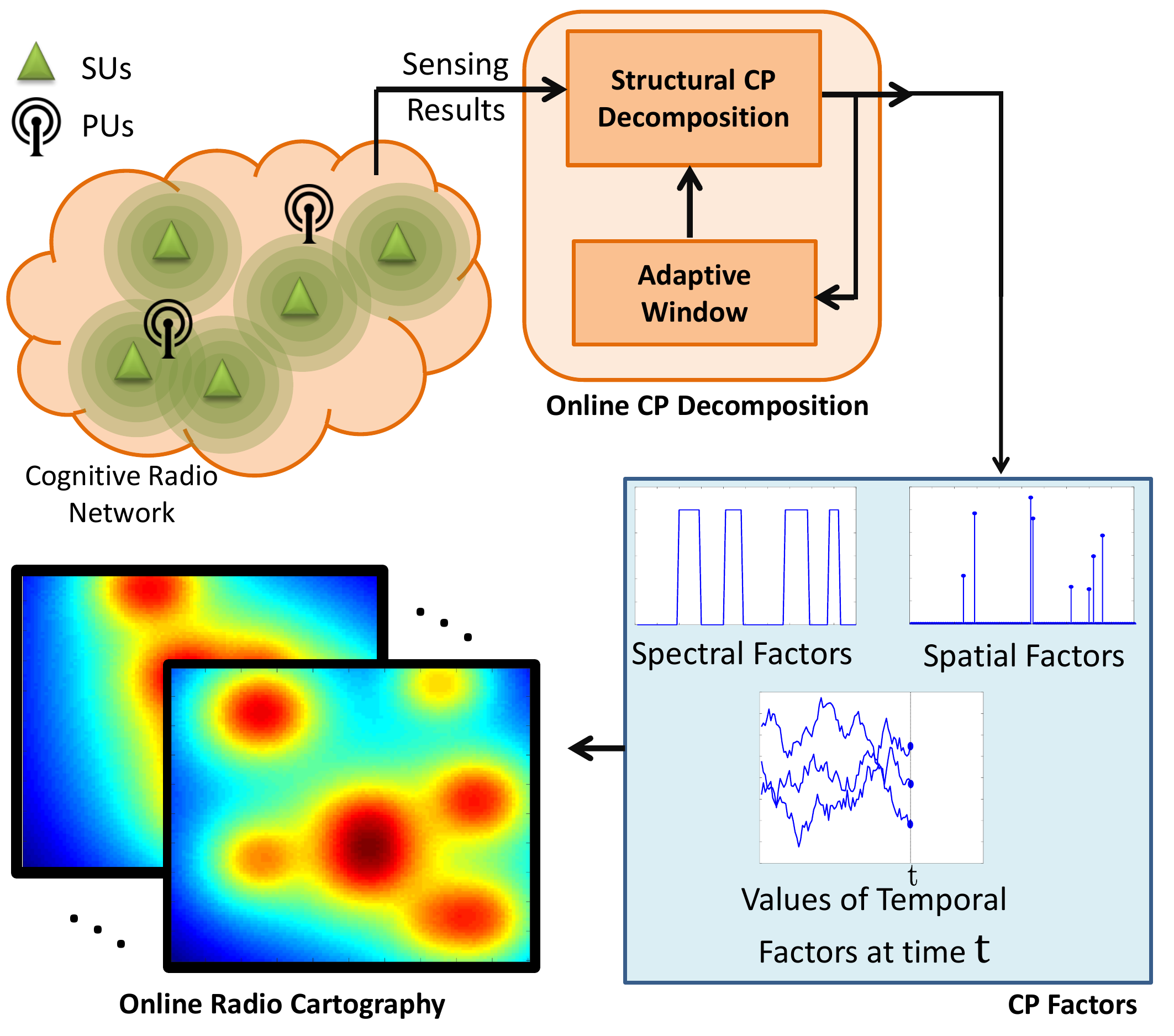}
\caption{\footnotesize{The block diagram of online implementation of the proposed structural CP for spectrum sensing. }}\label{fig:online_sensing}
\vspace{-3mm}
\end{figure}

The real-time requirements on dynamic radio cartography motivate us to extend the proposed framework to an online version. 
 A practical online implementation of the proposed framework should process received data from recent time slots and should not involve obsolete sensed data in order to infer CP factors. The CP factors of the online implementation enables us to reconstruct an online spectrum map. Fig. \ref{fig:online_sensing} shows the block diagram of the online implementation of our proposed algorithm. 

At time slot $t$, the recent $W(t)$ sensed spectrum measurements over $F$ frequency bins from $N$ SUs are aggregated in an $N\times F \times W(t)$ tensor. Here, $W(t)$ is an integer greater than $1$ and is called the temporal window at time slot $t$.

The sensed tensor at time $t$ is limited to the recent measured data in tensor $\underline{\boldsymbol{Y}}$. In the online implementation, the length of window should be optimized according to the dynamic of the system. To this aim, we employ the \emph{additive increase multiplicative decrease} model \cite{zhou2013study}. We propose the following criterion for updating the adaptive window,

\begin{equation}
\small
W(t) = 
     \begin{cases}
      W(t-1)+1 &\quad \text{if}\; \|\underline{\boldsymbol{Y}}-\underline{\boldsymbol{X}}\times_1\boldsymbol{\Gamma}\|_F\le J\\
      1+\floor{W(t-1)/2}{} &\quad\text{if}\; \|\underline{\boldsymbol{Y}}-\underline{\boldsymbol{X}}\times_1\boldsymbol{\Gamma}\|_F> J,\\
    \end{cases}
    \label{eq:window}
\end{equation}  
where, {\small $\|\underline{\boldsymbol{Y}}-\underline{\boldsymbol{X}}\times_1\boldsymbol{\Gamma}\|_F$ } is the metric to detect gross changes in the received dynamic spectrum and $J$ is the threshold to distinguish  that the received data is fit to the model or obsolete data must be removed in order to learn CP factors from more fresh measurements. The received data are organized in a tensor at time slot $t$ and the proposed structural CP decomposition is performed at each time slot. A computationally-efficient algorithm for a practical system can be performed using recent advances in online CP algorithms which is out of scope of this paper and we refer the interested readers to~\cite{zhou2016accelerating}.

\section{Experimental Results}
\label{sec:exper}
 We assume the system model in Sec. \ref{sec2}, where a small number of source points are propagating in a relatively narrow spectrum band. Let us construct the 3-way tensor, $\underline{\boldsymbol{Y}}$, in which each entry shows the power spectrum at each location, frequency, and time. We want to estimate the power propagation tensor $\underline{\boldsymbol{X}}$. The measurements of sensors are contaminated by AWGN with SNR of $+5$ dB in addition to perturbation with the shadowing Rayleigh channel with 6 tabs of multipath. $25$ candidate points are considered in the area of network.  
 Each sensor measures a superposition of sources signal that is multiplied by channel gains. Moreover, we consider $F=64$ frequency bins and $T=100$ time slots to generate the synthetic sensed tensor. 

\begin{figure}[b]
\centering
\vspace{-2mm}
\includegraphics[width=3in,height=1.85in]{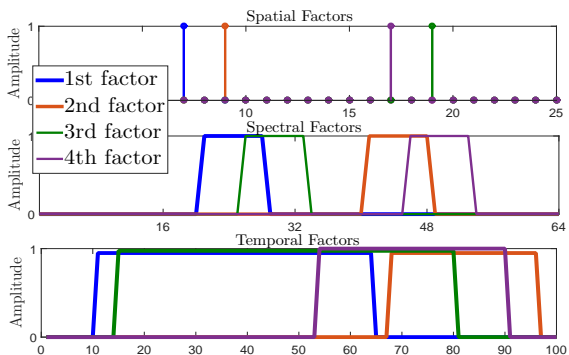}
\caption{\footnotesize{Ideal factors that synthesized data are generated accordingly.}}\label{fig:neat_factors}
\end{figure}

Fig. \ref{fig:neat_factors} represents the ideal factors of $\underline{\boldsymbol{X}}$ for our spectrum sensing synthetic data. This figure identifies a rank-4 tensor. Factors of a same rank-1 component are shown in the same color. The signature of propagation can be identified for every location in every frequency and at every time by interpreting these factors. For example, from the figure we can infer that there is propagation at location 9 in frequency bins 41 to 48 from time slot 68 to 97 (follow orange color in the patterns). This factors only are considered as the ground truth. We should note that the goal is to estimate $\underline{\boldsymbol{X}}$ given we only have access to contaminated and compressed data at sensors ($\underline{\boldsymbol{Y}}$).  

\begin{figure}[t]
\centering
\includegraphics[width=3 in,height=1.4in]{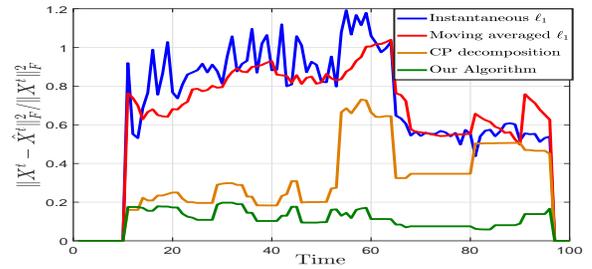}
\caption{\footnotesize{Spectrum estimation error for several algorithms.}}\label{fig:time_tens_mtx}
\vspace{-3mm}
\end{figure}

\begin{figure}
\centering
\includegraphics[width=2.6 in,height=1.45in]{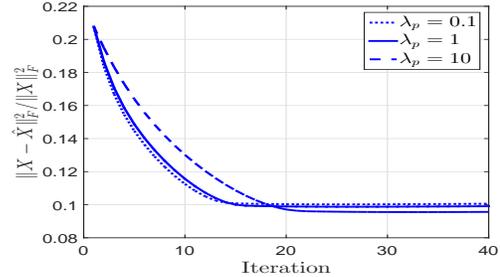}
\caption{\footnotesize{The effect of regularization parameter $\lambda_p$ on convergence.}}\label{fig:lambda_effect}
\vspace{-3mm}
\end{figure}

 The first simulation compares the performance of source localization using our proposed tensor-based algorithm and conventional  approaches.  
Fig. \ref{fig:time_tens_mtx} compares performance of tensor-based localization and the performance of localization using    the slices of the sensed tensor independently. Each slice of tensor is cast as a vector and an $\ell_1$ minimization problem is solved for estimating the power map using perturb compressive sensing algorithm as proposed in \cite{zhu2011sparsity}. Moreover,  the slice-based algorithm is applied on moving averaged slices of tensor. In each time slot 10 recent slices are averaged. The performance of CP decomposition is shown as the initialization of our algorithm. Each slice of the propagation tensor is compared with the ground truth and the normalized error is depicted. In this figure. $\boldsymbol{X}^t$ and $\boldsymbol{\hat{X}}^t$ are the ground truth and the estimated propagation at time $t$, respectively. As can be seen, our proposed algorithm outperforms the other methods in terms of normalized error of power map estimation. Since our proposed method is iterative, we need to study its convergence. The convergence behavior of our algorithm is exhibited in Fig. \ref{fig:lambda_effect}. As we increase $\lambda_p$, the algorithm converges slower, however, it convergence to a little bit more accurate solution. 

\begin{figure}[b]
\centering
\includegraphics[width=3in,height=2in]{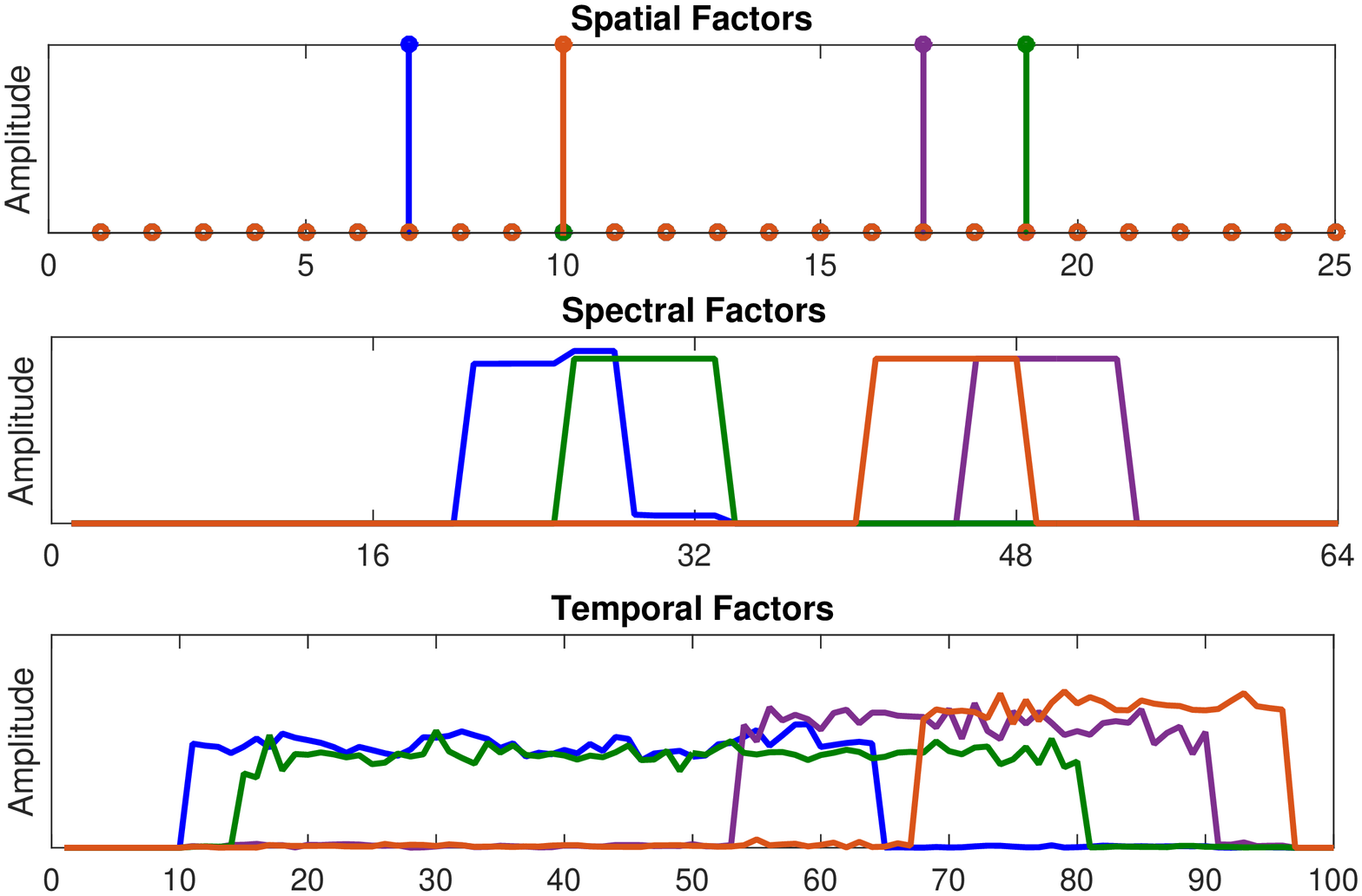}
\caption{\footnotesize{Reconstructed factors in simulation of Fig. \ref{fig:time_tens_mtx} in presence of multi-path fading and noise.}}\label{fig:cal_factors}
\end{figure}

Fig. \ref{fig:cal_factors} shows the obtained spatial, spectral and temporal factors using our method. As shown, the locations of PUs and their corresponding bandwidth occupancy are estimated accurately (compare it with the actual patterns in Fig.~\ref{fig:neat_factors}.   

\begin{figure}[t]
\centering
\begin{subfigure}[b]{0.15\textwidth}
\centering
\includegraphics[width=1in,height=0.85in]{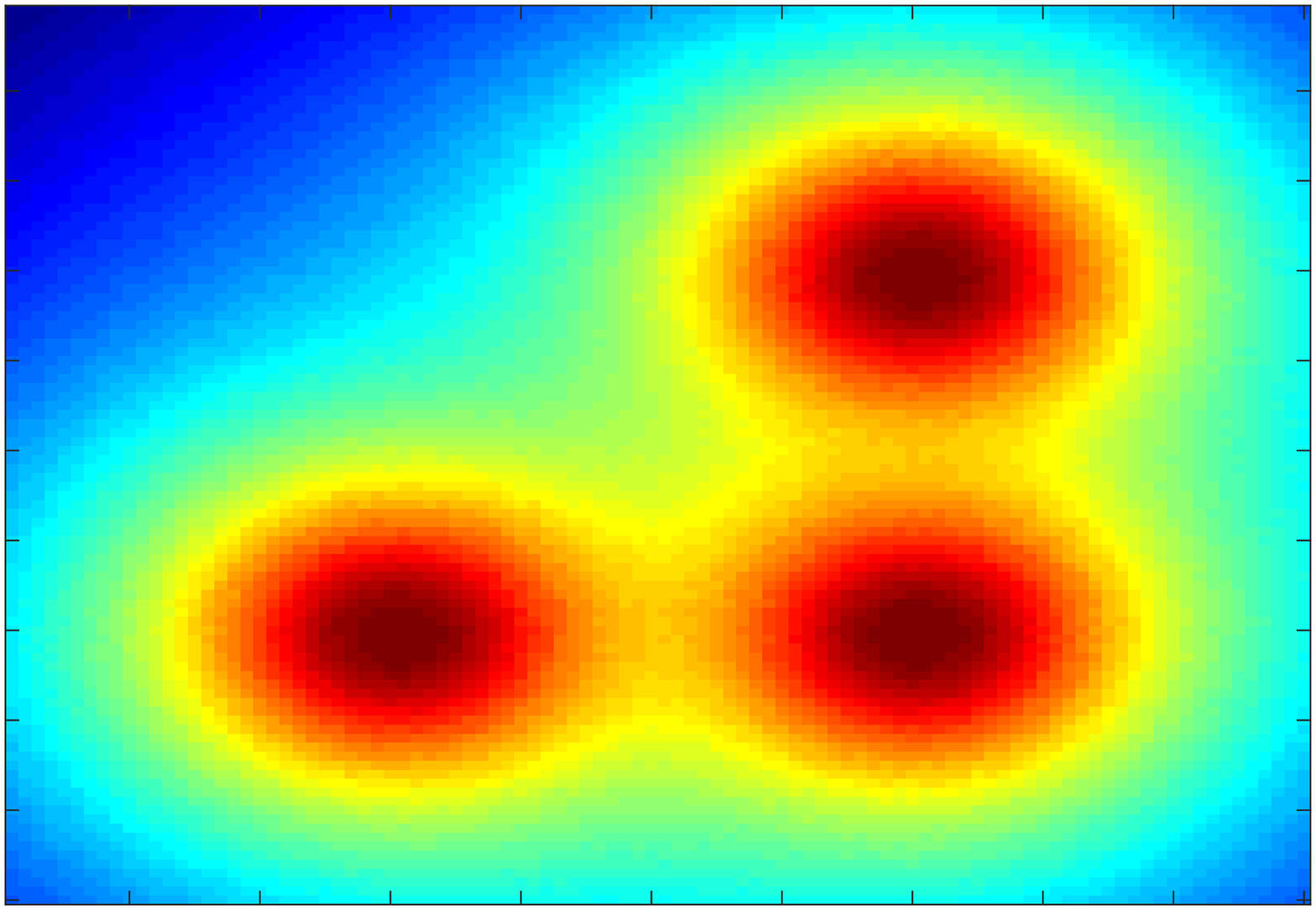}
\vspace{-1mm}
\caption{}
\label{fig:power_map:1}
\end{subfigure}
\begin{subfigure}[b]{0.15\textwidth}
\includegraphics[width=1in,height=0.85in]{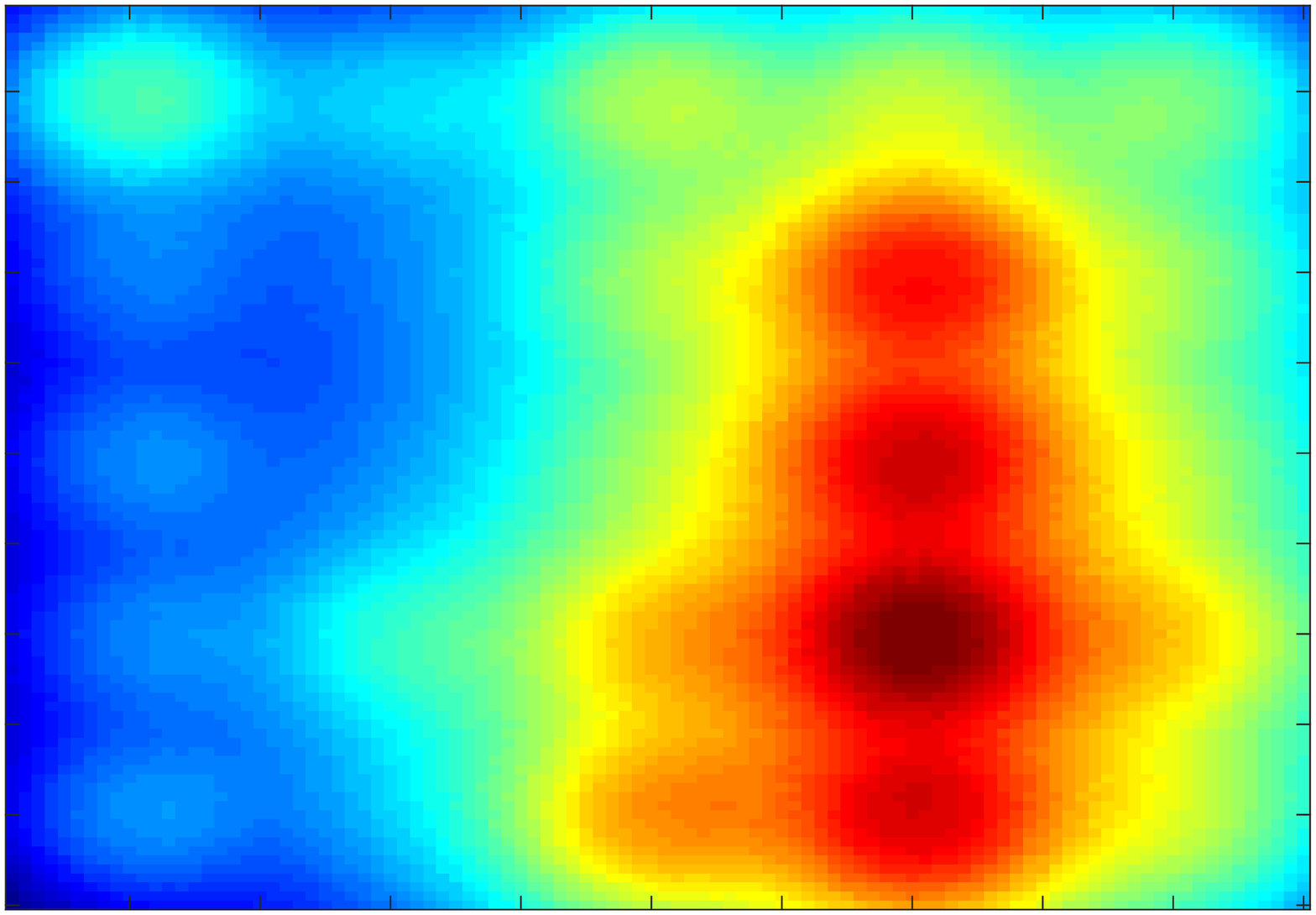}
\vspace{-1mm}
\caption{}
\label{fig:power_map:2}
\end{subfigure}
\begin{subfigure}[b]{0.15\textwidth}
\includegraphics[width=1in,height=0.85in]{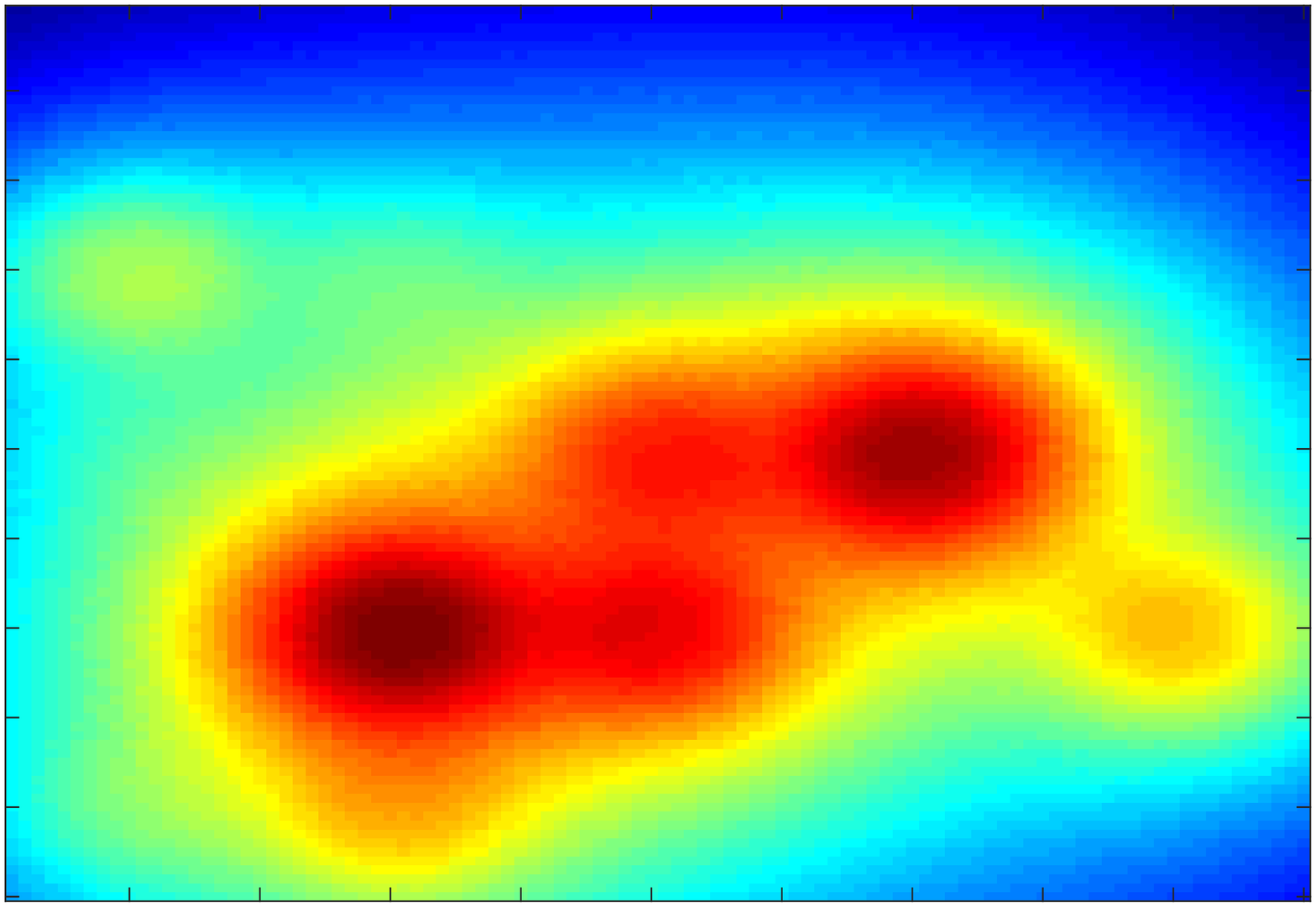}
\vspace{-1mm}
\caption{}
\label{fig:power_map:3}
\end{subfigure}
\begin{subfigure}[b]{0.15\textwidth}
\includegraphics[width=1in,height=0.85in]{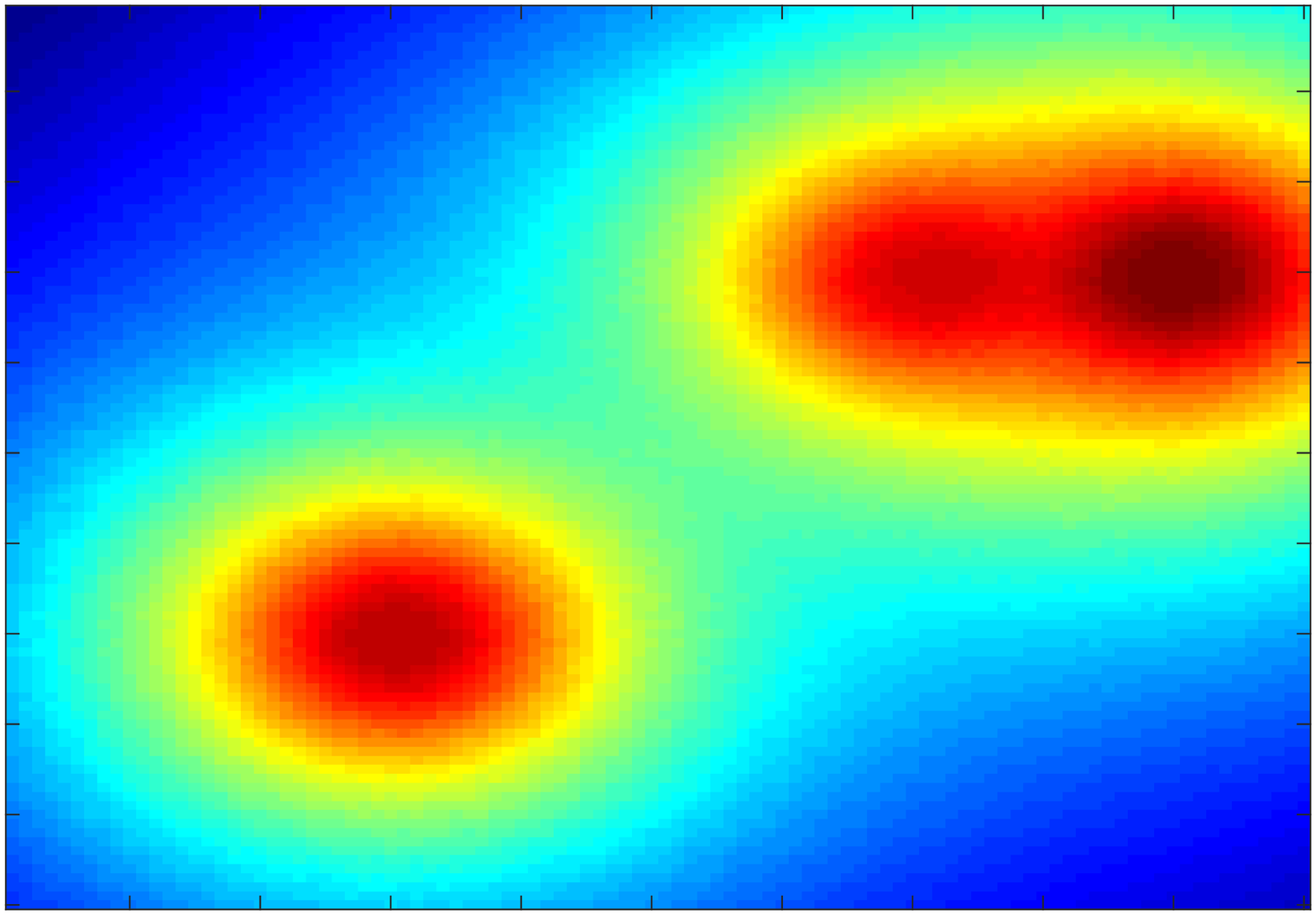}
\vspace{-1mm}
\caption{}
\label{fig:power_map:4}
\end{subfigure}
\begin{subfigure}[b]{0.15\textwidth}
\includegraphics[width=1in,height=0.85in]{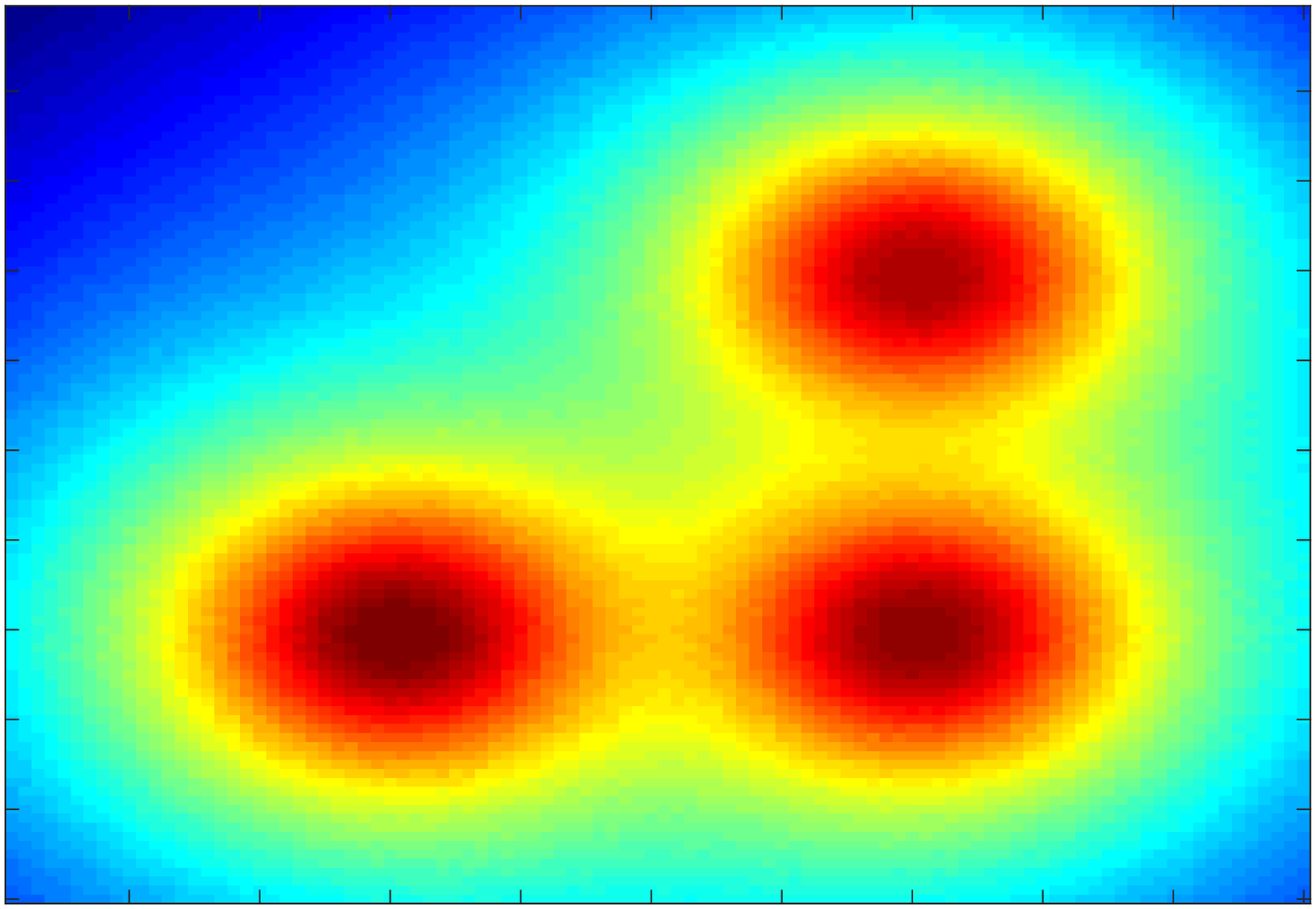}
\vspace{-1mm}
\caption{}
\label{fig:power_map:5}
\end{subfigure}
\vspace{-1mm}
\caption{\scriptsize {The spectral-aggregated power spectrum in time slot 60 for Fig.
\ref{fig:time_tens_mtx}. (a) The original power spectrum map. (b) The least squares solution. (c) The LASSO solution. (d) Extracted spectrum map using basic CP decomposition. (e) Extracted spectrum map using the proposed iterative algorithm.}}
\vspace{-1mm}
\label{fig:power_map}
\end{figure}

The next experiments exhibits the accuracy of our method for radio cartography. The spectrum maps are reconstructed using Eq. (\ref{eq:power_map}). Fig. \ref{fig:power_map:1} indicates the true spectrum map 
in time slot 60 which is extracted by the ideal factors of the tensor. This figure is the aggregated spectrum throughout all spectrum. In time slot 60 in Fig. \ref{fig:neat_factors} there are three active components. Locations 7, 17 and 19 are propagating in a narrow spectrum band. The obtained power map for all frequencies are aggregated in one map in this figure.  Fig. \ref{fig:power_map:2} shows the naive least squares solution and Fig. \ref{fig:power_map:3} shows $\ell_1$ regularized least squares solution known as LASSO. This solution is more accurate than the simple least squares solution, however, a dominant spurious term is visible in this solution. Fig. \ref{fig:power_map:4} exhibits the extracted solution using CP decomposition. There is only a weak spurious term in this solution which is alleviated by our iterative method which is shown in Fig. \ref{fig:power_map:5}.

The final experiment shows efficiency of the proposed online framework in Fig. \ref{fig:online_sensing} for a dynamic network. Consider a network with $25$ grid points and $15$ sensors in which $2$ PUs are active and they move each $50$ time slots. The proposed online framework detects a gross change in the dynamic of network and updates the length of processing window based on (\ref{eq:window}).

\section{Conclusion}
\label{sec:conc}
Tensor decomposition is a powerful tool for high dimensional data analysis. Spectrum data that are captured from a set of sensors in different frequencies and times are organized in a tensor and it is modeled by the compressed replica of a latent tensor. Decomposition of the compressed  tensor reveals for us factors of the latent tensor. Then, spatial, spectral and temporal activation pattern of propagating power are extracted. Scarce presence of PUs, their narrow band communication and smooth behavior of PUs over time suggest proper structures on CP factors. Moreover, uncertainty of the compression operator is modeled by a perturbation matrix and it is optimized in an iterative manner.    
\begin{figure}[t]
\centering
\includegraphics[width=3in,height=1.71in]{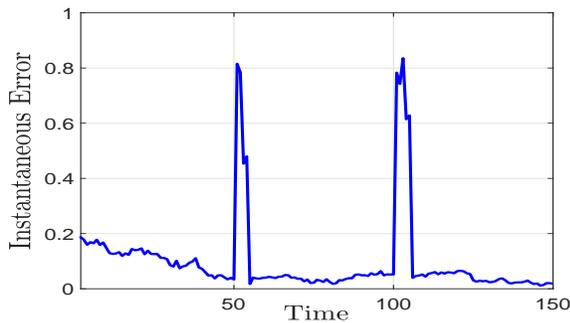}
\caption{\footnotesize{The performance of proposed online framework over time in terms of instantaneous error.}}\label{fig:online_simul}
\end{figure}

\footnotesize{
\balance
\bibliography{ref.bbl}
\bibliographystyle{IEEEtran}

}

\end{document}